\documentclass[manuscript]{aastex}

\usepackage{tikz}
\usepackage{amsmath}

\usepackage{graphicx}

\usepackage{pdflscape}

\slugcomment{The Astronomical Journal}

\shorttitle{The Trojan Color Conundrum}
\shortauthors{Jewitt}

\begin{document}

\title{The  Trojan Color Conundrum}


\author{David Jewitt$^{1,2}$
}

\affil{$^1$Department of Earth, Planetary and Space Sciences,
UCLA, 595 Charles Young Drive East, Los Angeles, CA 90095-1567\\
$^2$Department of Physics and Astronomy, University of California at Los Angeles, 430 Portola Plaza, Box 951547, Los Angeles, CA 90095-1547\\
}

\email{jewitt@ucla.edu}

\begin{abstract}
The Trojan asteroids  of Jupiter and Neptune are  likely to have been captured  from  original heliocentric orbits in the  dynamically excited (``hot'') population of the Kuiper belt.  However, it has long been known that the optical color distributions of the Jovian Trojans and the hot population are not alike.  This difference has been reconciled with the capture hypothesis by assuming that the Trojans were resurfaced (for example, by sublimation of near-surface volatiles) upon inward migration from the Kuiper belt (where blackbody temperatures are $\sim$40 K) to Jupiter's orbit ($\sim$125 K).  Here, we  examine the optical color distribution of the \textit{Neptunian} Trojans using a combination of new optical photometry  and published data. We find a  color distribution that is statistically indistinguishable from that of the Jovian Trojans but unlike any sub-population in the Kuiper belt.  This result is puzzling, because the Neptunian Trojans are very cold (blackbody temperature $\sim$50 K) and a thermal process acting to modify the surface colors at Neptune's distance would  also affect the Kuiper belt objects beyond, where the temperatures are nearly identical.   The distinctive color distributions of the Jovian and Neptunian Trojans thus present us with a conundrum:  they are very similar to each other, suggesting either capture from a common source or surface modification by a common process.  However, the color distributions differ from any plausible common source population, and there is no known modifying process that could operate equally at both  Jupiter and Neptune.  

\end{abstract}

\keywords{Kuiper belt: general---planets and satellites: dynamical evolution and stability}

\section{INTRODUCTION}
Jupiter's orbit is shared by so-called ``Trojan''  asteroids, which librate around the L4 and L5 Lagrangian points of the Sun-Jupiter system (see Slyusarev \& Belskaya 2014 for a recent review).  In most modern theories, the Trojans are thought to have been captured from initial heliocentric orbits, but the specific mechanism of capture remains unknown.  Primordial capture has been suggested (Marzari and Scholl 1998, Chiang and Lithwick 2005). However, simulations indicate that planetary migration would destabilize any primordially captured Trojans (Kortenkamp et a.~2004), and most models assume that the Trojans were captured stochastically during the clearing of the trans-Neptunian disk (Morbidelli et al.~2005, Lykawka et al.~2009, Parker 2015, Gomes and Nesvorny 2016).  The similarity between the size distribution of large Jovian Trojans and of Kuiper belt objects has been advanced as evidence for capture of the former from the latter (but, with complications, c.f.~Section \ref{otherevidence}).  However, compelling evidence for a connection is lacking.  The optical color distribution of the Jovian Trojans is weakly bimodal (Szabo et al.~2007) but, while they are red compared to most other objects in the asteroid belt (Grav et al.~2012, Chatelain et al.~2016), the Trojans are completely lacking in the ultrared surfaces (B-R $>$ 1.6) that are a distinctive feature of the Kuiper belt and Centaur populations (Jewitt 2002, 2015, Lacerda et al.~2014).   

Neptune  also has Trojans (Sheppard and Trujillo 2006, hereafter ST06). In this paper, we combine new measurements of the optical colors of six Neptunian Trojans with measurements from the published literature (Parker et al.~2013, ST06) to define their properties as a group. Our objective is to compare the color distributions of the two Trojan populations, both with each other and with  other solar system groups, in order to search for hints about possible relationships.  

\section{OBSERVATIONS}
We used the Keck 10 m diameter telescope atop Mauna Kea (altitude 4200 m) with the Low Resolution Imaging Spectrometer (LRIS: Oke et al.~1998) in order to obtain optical photometry of the Neptunian Trojans.  LRIS possesses independent blue and red channels separated by a dichroic filter.  We used the ``460'' dichroic which has 50\% peak transmission at 4900\AA~wavelength, and  a broadband B filter on the blue side.  The B filter has central wavelength $\lambda_C$ = 4370\AA~ and Full Width at Half Maximum (FWHM) = 878\AA.  On the red side, we alternated between broadband V ($\lambda_C$ = 5473\AA, FWHM = 948\AA) and R ($\lambda_C$ = 6417\AA, FWHM = 1185\AA) filters.  Typical integration times were $\sim$300 s, during which the telescope was tracked to follow the non-sidereal motions of the Trojans while simultaneously guiding on a nearby field star.  The identities of the Trojans, which are faint enough to be confused with background Kuiper belt objects, were confirmed by their expected positions and non-sidereal rates.  Two Trojans (2004 KV18; Horner and Lykawka 2012 and 2012 UW177; Alexandersen et al.~2016) are thought, on the basis of numerical integrations of the equations of motion, to be temporary captures from the Centaur population, and were not observed.  Photometric calibration of the data was secured using observations of  standard stars selected to have sun-like colors from the list by Landolt (1992).  The seeing was typically $\sim$0.8\arcsec~FWHM.  Repeated measurements of a given field and data from simultaneous operation of the CFHT Sky Probe monitor showed that each night was photometrically stable to $\pm$0.01 magnitude.  A journal of observations is given in Table (\ref{geometry}).

We flattened the data using bias frames and flat-field images obtained from a uniformly illuminated patch inside the Keck dome. Aperture photometry was used to measure the brightness of the Trojans in each filter.  We selected apertures  based on the seeing, settling for most objects on projected radius 2.8\arcsec~with sky subtraction from the median pixel value in a contiguous annulus extending to 5.6\arcsec.  A few images in which the target Trojans appeared blended with field stars or galaxies, or were irretrievably compromised by image blemishes, were omitted from further consideration.  
 
The photometric results are summarized in Table (\ref{photometry}).  Absolute magnitudes, $H_R$, were computed from the apparent magnitudes using the inverse square law and the HG formalism (Bowell et al.~1989) with assumed phase angle parameter $G$ = 0.15, as is appropriate for dark surfaces.  The phase darkening coefficients are unmeasured, however, introducing some uncertainty into $H_R$, particularly if the Trojans should show significant opposition surge (although available evidence from the Jovian Trojans indicates that they do not; Shevchenko et al.~2012).  Values of $H_R$ are quoted only to one decimal place in recognition of this phase function uncertainty.  For reference, the smallest and largest effective radii computed from the data in Table (\ref{photometry}), assuming geometric albedo $p_V$ = 0.06, are 43 km (2005 TN53) and 130 km (2013 KY18).

\section{DISCUSSION}

The Kuiper belt objects display a wide range of optical colors, likely indicating a wide range of surface compositions (Luu and Jewitt 1996, Tegler and Romanishin 2000, Jewitt and Luu 2001, Jewitt 2002, Hainaut et al.~2012, Sheppard 2010, 2012, Lacerda et al.~2014, Peixinho et al.~2015).  A significant fraction of the Kuiper belt objects, notably but not exclusively those in the low inclination cold-classical population, are ultrared (defined by having normalized optical reflectivity gradients $S' \ge$ 25\%/1000\AA, corresponding to B-R $>$ 1.6, Jewitt 2002).  The material responsible for the ultrared color is not known with certainty, but is commonly identified with irradiated complex organic matter (Cruikshank et al.~1998, Jewitt 2002, Dalle-Ore et al.~2015, Wong and Brown 2017).  The working picture is of a meter-thick shell of hydrogen-depleted complex organics, processed by exposure to the cosmic ray flux and underlain by unirradiated matter with, presumably, different optical properties.  Apart from the KBOs, only the Centaurs (themselves recent escapees from the Kuiper belt) show ultrared surfaces.

We list in Table (\ref{colors})  the colors of  several dynamically defined sub-populations in the Kuiper belt, and of the Centaurs, all taken from Peixinho et al.~(2015).    For each group, we list the number of objects, and the mean  and median colors for each of  B-V, V-R and B-R.  The listed uncertainties are the formal $\pm$1$\sigma$ errors on the means, and do not reflect possible systematic errors in the photometric calibration.  We use the unweighted means to avoid giving undue weight to the brightest, most easily measured objects in each population.   To estimate the systematic errors, we compared measurements   compiled from  independent sources (namely, Table 10 of Jewitt 2015 and the ``MBOSS'' compilation of Hainaut et al.~2012), with those listed in the Table.  The root-mean-square differences between the colors in this reference and the colors in Table (\ref{colors}) are $\sim$1.5$\sigma$ in B-V and $\sim$0.6$\sigma$ in V-R, showing that the systematic errors, while slightly larger in B-V than in V-R, are in both cases comparable to the random errors.   Presumably, the systematic errors originate with the use of different  filters, detectors  and calibration stars  by different researchers and also, especially in the case of the cold-classical Kuiper belt objects, from small differences in the adopted definitions of the sub-populations). Most importantly, the random and systematic errors combined are small compared to the color differences between populations in  Table (\ref{colors}).  

The Neptunian Trojans (Table \ref{photometry})  occupy a range of absolute magnitudes,  6.3 $\le H_R \le$ 8.6, which is well matched to the Kuiper belt objects (because they are at similar distances).  On the other hand, the  observed Jovian Trojans are much closer and, for a given apparent magnitude, intrinsically about 10 times smaller than the Neptunian Trojans and Kuiper belt objects, potentially introducing a size dependent bias.  We believe this size bias to be negligible given that the range of colors in the Kuiper belt does not depend strongly on $H_R$  and there is no hint, for example, that small KBOs are preferentially blue (c.f.~Figure 6 of Peixinho et al.~2012, Wong and Brown 2017).  We  also considered the possibility of bias caused by wavelength dependence of the scattering phase function (``phase reddening'').  This might affect the Jovian Trojans systematically, because they  attain  larger phase angles  (up to $\sim$12\degr) than do the more distant bodies of the outer solar system  in  Table (\ref{colors}).  Fortunately, available measurements of the phase reddening coefficient are consistent with zero (Chatelain et al.~2016) and so we believe that this effect is also negligible.    

\begin{figure}
\epsscale{0.85}
\plotone{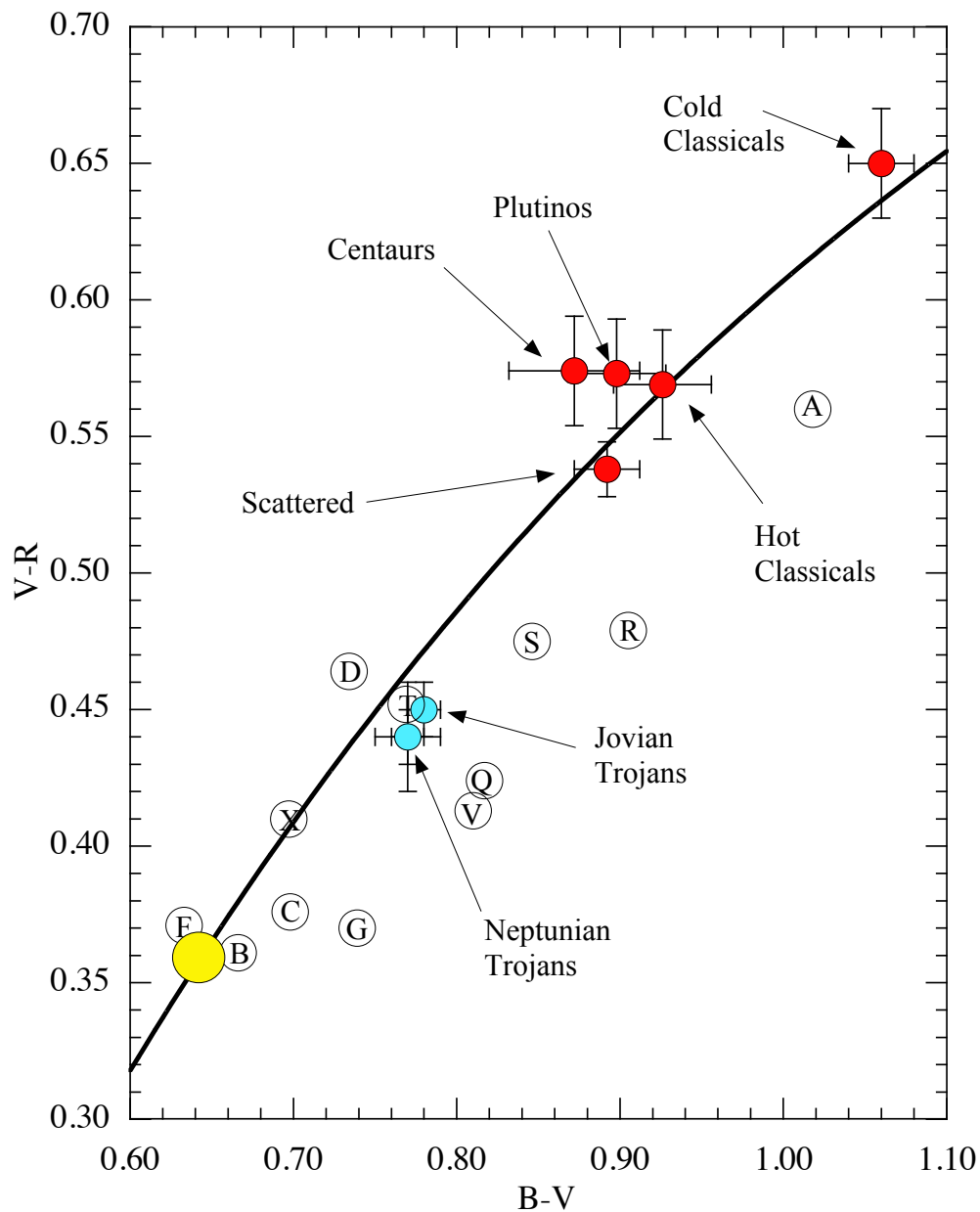}
\caption{Color-color diagram showing average B-V vs.~V-R colors (with $\pm$1$\sigma$ errors on the means) for the Trojans (blue circles) and other populations (red circles), as labeled.  Letters mark the approximate locations of different asteroid spectral types, from Dandy et al.~(2003). The color of the Sun is marked by a yellow circle.  The line shows the locus of points having linear reflectivity spectra.
\label{colorcolor}}
\end{figure}

\begin{figure}
\epsscale{0.8}
\plotone{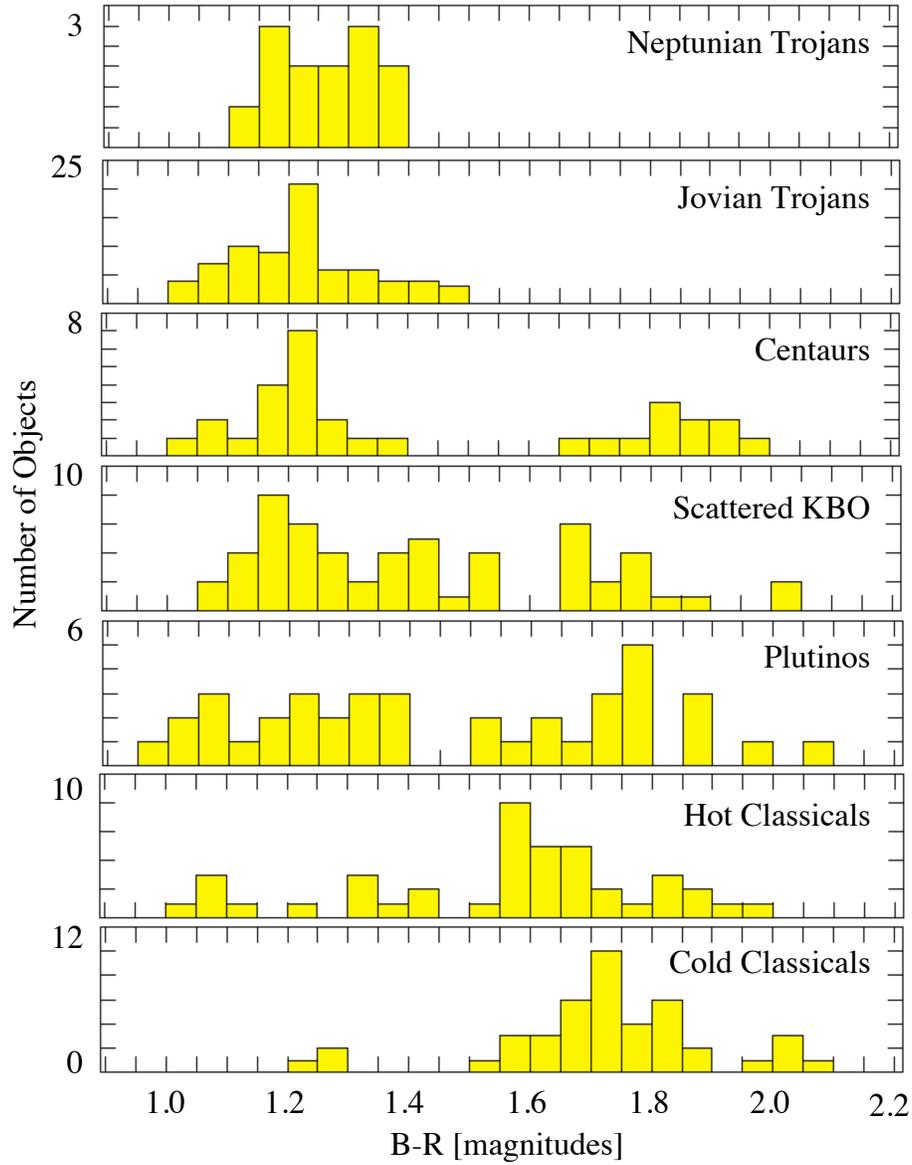}
\caption{Histograms of B-R for each of the measured populations.  The numbers of objects in each sample are listed.  \label{histo}}
\end{figure}

The resulting mean colors of the Neptunian Trojans,  B-V = 0.77$\pm$0.01, V-R = 0.44$\pm$0.01, B-R = 1.20$\pm$0.03,  are redder than the Sun, for which (B-V)$_{\odot}$ = 0.64$\pm$0.02, (V-R)$_{\odot}$ = 0.35$\pm$0.01, (B-R)$_{\odot}$ = 0.99$\pm$0.02 (Holmberg et al.~2006).  The color data are  plotted in Figure (\ref{colorcolor}) and shown as histograms in Figure (\ref{histo}).    It is immediately clear from Figures (\ref{colorcolor}), (\ref{histo}) and Table (\ref{colors}) that the Jovian and Neptunian Trojan colors are alike, but very different from the mean colors of the other small-body populations.   This confirms an earlier report based on photometry of six objects, to the effect that the Neptunian Trojans are distinguished by the absence of ultrared members (Sheppard 2012). 

\begin{figure}
\epsscale{1.0}
\plotone{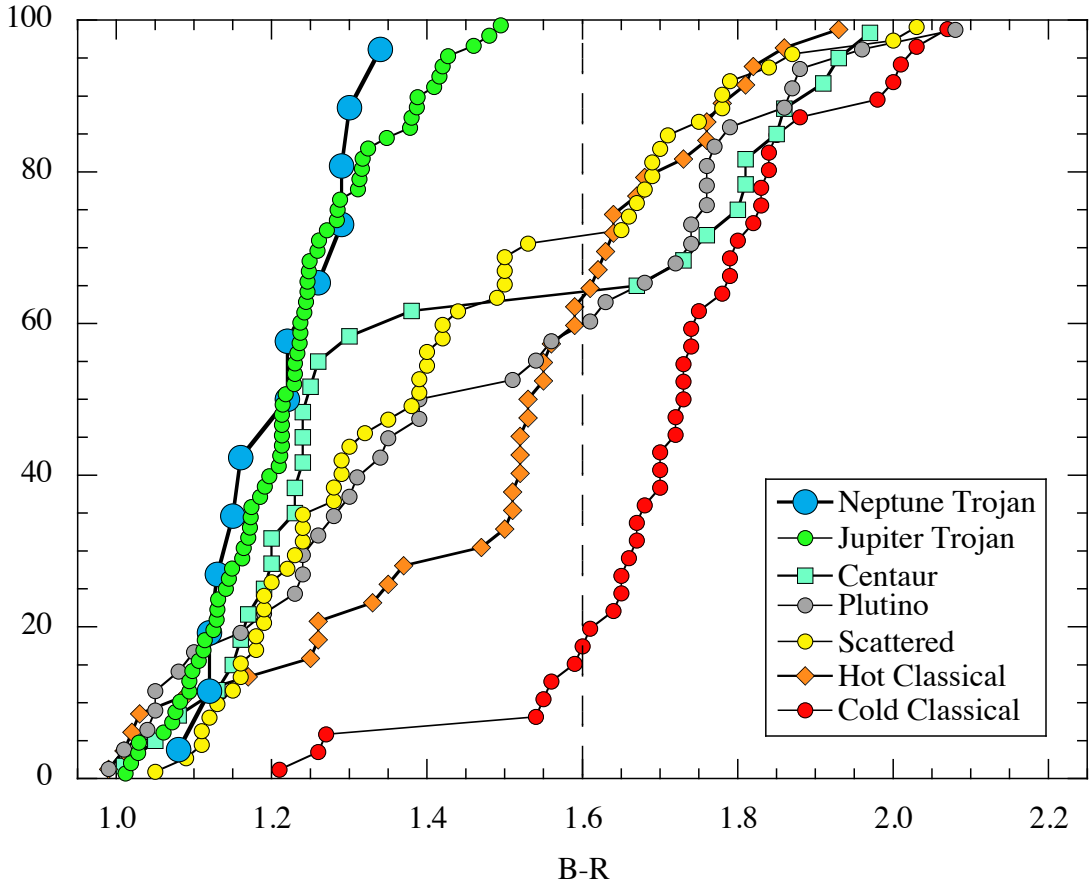}
\caption{Cumulative distributions of the B-R color index for the Neptunian Trojans and outer solar system populations discussed in the text.  The dashed vertical line separates ultrared objects (to the right)  from the others.
\label{cumu}}
\end{figure}

To explore these similarities and differences, we compare the cumulative color distributions  in Figure (\ref{cumu}).  The Figure shows that the principal differences between the populations lie in the fraction of ultrared objects.   We use two statistical tests to quantify the differences evident in the figure.  The  Kolmogorov-Smirnov (K-S) test is essentially a measure of the maximum difference between any two cumulative curves.  Specifically, the K-S test provides a non-parametric estimate of the null hypothesis that any two color distributions could be drawn by chance from the same parent population.  The Anderson-Darling (1954) test is similar to the K-S test but is more sensitive to differences at the tails of the distribution.  We use the B-R color index as our metric, motivated by the observation that the reflectivity spectra of outer solar system bodies  are linear with respect to wavelength across the optical spectrum (Jewitt 2015).  The results are summarized for both tests in Table (\ref{KS}).  

The statistical tests  in Table (\ref{KS}) reinforce what is obvious to the eye in the histograms of Figure (\ref{histo}), namely, that the Jovian and Neptunian Trojan color distributions are statistically consistent with being drawn from a common parent population but  that they are unlike other outer solar system populations (Figures \ref{colorcolor}, \ref{histo} and \ref{cumu}).   We further duplicated these conclusions by separately conducting the entire analysis using the dataset from Hainaut et al.~(2012).

\subsection{Trojans and the Kuiper Belt}

In the Nice and related models, dynamical instability of an initially massive Kuiper belt feeds numerous niche populations, including the Trojans, with escaped Kuiper belt objects.  In these models, the surviving counterparts to the escaped objects are members of the dynamically excited, so-called ``hot'' populations, including the hot-classical objects, the scattered Kuiper belt objects, and the resonant objects.   It is thus natural to expect that the colors of the Trojans should resemble those of the hot populations, but Table (\ref{colors}) shows that they do not.   A convenient way to describe this is in terms of the fraction of ultrared objects, defined as those having B-R $>$ 1.6 (Jewitt 2002), in each population.  Figure (\ref{cumu}) shows that the cold-classicals are about 4/5ths ultrared, while the hot-classical, scattered and 3:2 resonant populations are together about 1/3rd ultrared. The Trojans contain no ultrared objects.

The  conundrum  raised by the data is that the two Trojan color distributions closely match each other (the populations have identical average colors, within the uncertainties of measurement, c.f.~Table \ref{colors}), but they do not resemble the suspected Kuiper belt source population from which they were captured. We see two comparably unsatisfactory solutions to the Trojan color conundrum.  

Solution 1: Kuiper Belt is Not the Source.   The simplest interpretation  is that the Trojans of Jupiter and Neptune lack ultrared matter because they did not form in the Kuiper belt and have no relation to the modern-day hot population.   The question then becomes ``where did they form?''.  Other formation locations have indeed been proposed.  For example, some models posit capture from source regions local to each planet.   These  include the pull-down capture model (Marzari and Scholl 1998) in which the rising mass of a growing planet stabilizes objects already near the leading and trailing Lagrange points, and the in-situ growth model (Chiang and Lithwick 2005) in which the protoplanetary disk is so dynamically dense and cold that Trojans accumulate in-place.  However, local capture is also unsatisfactory given the vast separation between Jupiter and Neptune and the fact that strong compositional and color gradients exist between the inner and outer solar system (Jewitt 2002, 2015).  In local capture scenarios the close color similarity between the Trojans of Jupiter and Neptune could only be regarded as a coincidence.  

Solution 2: Surface Evolution. The optical properties of the Trojans could have been modified thermally, in response to  their inward displacement from the Kuiper belt  (Luu and Jewitt 1996, Jewitt 2002).    This is very plausible  at Jupiter (5 AU), where the isothermal blackbody temperature, $T_{BB}$ = 125 K, is much higher than in the Kuiper belt (40 AU and 45 K). Sublimation (and crystallization) of embedded ices would naturally and rapidly lead to the burial of an ultrared surface layer via the deposition of a mantle of fallback material (Jewitt 2002).   Support for this scenario comes from observations of the Centaurs.  Distant Centaurs, with perihelia $q \gtrsim$ 10 AU, exhibit a wide range of colors consistent with their extraction from the hot component of the Kuiper belt  (see Figure \ref{histo}, Jewitt 2015).  However, at smaller perihelion distances, the red-surfaced Centaurs are systematically depleted relative to their abundance at larger distances and, once captured as Jupiter family comets, all the ultrared matter is gone (Jewitt 2009, 2015).   Comet-like activity also begins in Centaurs with $q \sim$ 10 AU,  consistent with the distance at which exposed amorphous ice can first crystallize (Jewitt 2009, 2015) but smaller than the critical distance for the selective sublimation of H$_2$S proposed by Wong and Brown (2017).  Thermal effects on Kuiper belt objects displaced from 40 AU to 5 AU are to be expected even prior to their putative capture as Trojans.

On the other hand, these thermal processes cannot operate at Neptune's distance, where the isothermal blackbody temperature, $T_{BB}$ = 50 K, is too low for common volatiles to sublimate or even for amorphous ice to crystallize.   Moreover,  temperatures at 30 AU are barely different from temperatures in the Kuiper belt beyond: any thermal process  operating to destroy ultrared matter at the distance (and temperature) of Neptune would also  operate in the Kuiper belt just beyond it.  Therefore, thermal processes cannot account for the similarity between the Jovian and Neptunian color distributions, or their difference from the Kuiper belt populations.  An exception to this conclusion could arise if the Neptunian Trojans were scattered into orbits with perihelia $\lesssim$10 AU prior to capture, but this is a possibility for which we are aware of no  evidence from dynamical simulations.

What about non-thermal processes?  Collisions offer the most obvious such process.  If the  Trojans experience more intense collisional processing than objects in the Kuiper belt   then perhaps the ultrared surfaces could have been preferentially destroyed.  However, we are unaware of existing evidence for particularly intense collisional processing of the Trojans, which would have to occur on a short timescale in order to prevent the regrowth of the irradiation mantle after capture into 1:1 resonance.  The collisional rate in the Trojans swarms is dominated by small bodies which, in existing magnitude-limited surveys, remain essentally unobserved.  Thus, it is technically possible that the Trojans suffer collisional resurfacing at a disproportionately high rate but, in the absence of data, such an explanation would seem, at best, to be highly contrived.  

\subsection{Other Evidence}
\label{otherevidence}
Another way to compare the Trojans with the Kuiper belt is through their size distributions, since the dynamics of capture into 1:1 resonance are presumably size-independent.  
It is reasonable to expect that the size distribution of the Trojans should reflect the size distribution of  objects in the population from which they were captured.  The measurable proxy for size is $H$, the absolute magnitude (apparent magnitude reduced to unit heliocentric and geocentric distances and 0\degr~phase angle).  The $H$ distributions can be fitted by broken power-laws, with different slopes above and below a critical ``break magnitude''.   In objects fainter (smaller) than the break, the slope (which, for the Jovian Trojans is $\alpha$ = 0.40$\pm$0.05, as established by many workers from Jewitt et al.~2000 to Yoshida and Terai 2017) is produced collisionally and has little to do with the size distribution of the objects when they were formed. For the larger objects brighter than the break magnitude, disruptive collisions are rare and objects are presumed to preserve their original dimensions.   Therefore, the most useful comparison to be made is between the distributions of the bright (large) Trojans and the bright (large) hot-classical Kuiper belt objects that are purported to represent the population from which the Trojans were captured.  

The large Trojan power-law index has been repeatedly measured and found to be steep:  $\alpha$ = 0.9$\pm$0.2 (Jewitt et al.~2000),  $\alpha$ = 1.0$\pm$0.2 (Fraser et al.~2014), and $\alpha$ = 0.91$_{-0.16}^{+0.19}$ (Wong and Brown 2017).  Morbidelli et al.~(2009) found that the cold-classical objects (for which they obtained $\alpha$ = 1.1) are much more like the large Trojans than are the  hot-classicals (for which they reported a much flatter distribution  with $\alpha$ = 0.65).  This is the exact opposite of the result expected if the Trojans were captured from the hot population, and capture from the cold (dynamically undisturbed) population makes no sense.  Fraser et al.~(2014) re-made the comparison but included a size-dependent albedo to convert from absolute magnitude to size.  This  allowed them to reach the opposite result, namely that the  large cold-classicals have a distribution that is too steep ($\alpha$ = 1.5$_{-0.2}^{+0.4}$) to fit the Trojans but that the  hot-classical objects (with $\alpha$ = 0.87$_{-0.2}^{+0.07}$) are very similar.  

Two weaknesses remain in the size distribution comparison.  First, the size ranges of the  Trojans and measured Kuiper belt populations barely overlap.  The break magnitude for the hot-classicals is $H_{r'}^b$ = 7.7$_{-0.5}^{+1.0}$ (Fraser et al.~2014), corresponding to $H_V^b$ = 7.9$_{-0.5}^{+1.0}$ (assuming V-r' = 0.2, Fraser et al.~2008, Jewitt 2005).  Only one Trojan, 624 Hektor (with $H_V =$ 7.3), is brighter than $H_V^b$.

Second, there are very few  objects brighter than the Trojan break magnitude and the difference between the break magnitude and the brightest object is also very small, limiting the precision with which $\alpha$ can be determined.  For example,  the Trojan break magnitude $H_{r'}^b$ = 8.4$_{-0.1}^{+0.2}$ (Fraser et al.~2014) corresponds to  $H_V^b$ = 8.6$_{-0.1}^{+0.2}$.  The JPL Horizons catalog lists only 10 Trojans with $H_V <$ 8.6, providing a rather meagre sample with which to fit the absolute magnitude distribution.   Furthermore, the 1.3 magnitude difference between $H_V^b$ and the brightest Trojan corresponds to objects in the very small diameter ratio 1.8:1.  As a result, the best-fit power law index has considerable formal uncertainty, $\alpha$ = 0.9$\pm$0.2 (references cited above).   The size distribution comparison is consistent with capture from the hot component of the Kuiper belt, but it is hardly a convincing proof.

The angular momenta of individual objects should not be changed by their capture into the 1:1 resonance, suggesting a simple test of the hypothesis that Trojans are captured Kuiper belt objects. Specifically, if the Trojans were captured from the Kuiper belt, then their rotation distributions should be the same\footnote{However, the test should be applied to objects larger than a critical radius, $a_c$,  since smaller objects are potentially influenced by YORP radiation torques.  The YORP timescale scaled from measurements of main-belt asteroids is $\tau_Y \sim K_Y a^2 r_H^2$, where $a$ is the radius in km, $r_H$ is in AU and $K_Y \sim$ 1 Myr is a constant.  Setting $\tau_Y = 4.5\times 10^9$ yr and $r_H$ = 5 AU for the Jovian Trojans, we solve to find $a_c \sim$ 13 km. In the Kuiper belt with $r_H$ = 40 AU, $a_c \sim$ 2 km.}.    The mean angular rates (assuming double-peaked lightcurves) of Kuiper belt objects have been estimated at 3.1 day$^{-1}$ (72 objects; Duffard et al.~2009), 2.0$\pm$0.2 day$^{-1}$ (15 objects from Table 3 of Benecchi and Sheppard 2013) and 2.8$\pm$0.2 day$^{-1}$ (29 objects; Thirouin et al.~2014).  The mean rotation rate in the Jovian Trojans  is  smaller at 1.7$\pm$0.2 day$^{-1}$ (Szabo et al.~2017), with 20\% of the 56 Trojans in their sample having very slow rotation ($< 0.5$ day$^{-1}$).  However, while the Jovian Trojan and Kuiper belt  mean rotation rates are formally not equal, it is too early to conclude that the difference is real.  The comparison suffers from some of the same weaknesses that afflict the size distribution comparison.  For example, there is a  difference in the sizes of the objects  (mean $H \sim$ 11.4 for the Trojans studied by Szabo et al.~vs.~$H \sim$ 6.0 for the Kuiper belt objects, corresponding to about an order of magnitude in size).   Observational bias (e.g.~the greater difficulty in securing large-aperture telescope time sufficient to determine very long periods in the faint Kuiper belt objects) likely also plays a role.  The same case can be made for comparative measures of the shape distribution of large objects; we do not yet possess the necessary data. Still, there is reason to hope that these biases can be addressed in the not too distant future using systematic observations from all-sky surveys (e.g.~Pan-STARRS or the Large Synoptic Survey Telescope).

In summary, the color evidence does not support the hypothesis that the Jovian and Neptunian Trojans were captured from the hot population, the size distributions (for the few objects larger than the Trojan  break magnitude) are consistent with but do not convincingly establish this origin for the Jovian Trojans, while  comparisons based on the respective rotation period distributions are premature.

\subsection{Trojans and Centaurs}

The orbits of Trojans are weakly stable and some can escape from the Lagrangian clouds during the lifetime of the solar system.  Horner and Lykawka (2010) concluded that ``....the Trojans can contribute a significant proportion of the Centaur population, and may even be the dominant source reservoir''.  If this were true, there should be no ultrared Centaurs (because there are no ultrared Trojans) whereas, in fact, about 1/3rd of Centaurs are ultrared (Table \ref{KS} and Figures \ref{colorcolor} and \ref{histo}).   By the K-S test, there is a 3\% likelihood that the Neptunian Trojans and the Centaur colors are drawn from the same parent population, and a 0.1\% chance that the Jovian Trojans and the Centaurs are. While the possibility that escaped Trojans contribute to the Centaur flux cannot be excluded, the colors show that they are  not the dominant source of the Centaurs.

\section{SUMMARY}

We determine the average optical colors of Neptunian Trojans  and compare them with the Jovian Trojans and with potentially related source populations in the outer solar system.  We find that 

\begin{enumerate}

\item The  optical color distributions of the Jovian and Neptunian Trojans are  indistinguishable from each other, but they are statistically different from the Kuiper belt populations from which capture has been suggested. 

\item If the Jovian Trojans were captured from the Kuiper belt, then their less red colors could be explained by temperature-dependent resurfacing due to volatile loss, as is observed in the Centaurs. In contrast, the Neptunian Trojans are too cold (and too similar in location and temperature to the Kuiper belt itself) for thermal effects to play a role. The observed equality of the color distributions of the Jovian and Neptunian must have another cause.

\item The  Trojan color distributions  are additionally distinct from the Centaur distribution, negating the hypothesis (Horner and Lykowka 2010) that escaped Trojans might dominate the Centaur population.

\end{enumerate}

\acknowledgments
I thank Nuno Peixinho for an easy-to-read  version of his color datafile, Jing Li and Pedro Lacerda for help with Mathematica, and Wesley Fraser and the ``Statistics editor'' for their comments. NASA provided support for some of the observations via its Solar System Observations program.  The data presented herein were obtained at the W. M. Keck Observatory, which is operated as a scientific partnership among the California Institute of Technology, the University of California and NASA. The Observatory was made possible by the generous financial support of the W. M. Keck Foundation.



{\it Facilities:}  \facility{Keck}.

\clearpage


\clearpage

\begin{deluxetable}{llcrrrccccr}
\tablecaption{Observing Geometry 
\label{geometry}}
\tablewidth{0pt}
\tablehead{ \colhead{Object} & \colhead{UT Date}  & \colhead{$r_H$\tablenotemark{a}} & \colhead{$\Delta$\tablenotemark{b}} & \colhead{$\alpha$\tablenotemark{c}}   }
\startdata

2014 QO441  &  	2016 Aug 03 & 33.225 & 33.120 & 1.7 \\
2011 SO277  &  	2016 Aug 03 & 30.480 & 30.234 & 1.9  \\
2013 KY18    &  	2016 Aug 04 &  30.318 & 29.748 & 1.6 \\
2011 WG157  &  	2016 Aug 04 &  30.766 & 30.967 & 1.8 \\
2010 TS191   &  	2016 Aug 04 &  28.702 & 28.858 & 2.0 \\
2010 TT191  &  	2016 Aug 04 &  32.146 & 32.561 & 1.6 \\

\enddata


\tablenotetext{a}{Heliocentric distance, in AU}
\tablenotetext{b}{Geocentric distance, in AU}
\tablenotetext{c}{Phase angle, in degrees}

\end{deluxetable}

\clearpage

\begin{deluxetable}{clccrrclccr}
\tabletypesize{\scriptsize}
\tablecaption{Photometry 
\label{photometry}}
\tablewidth{0pt}
\tablehead{\colhead{Object Name} & \colhead{UT Date\tablenotemark{a}} & \colhead{R\tablenotemark{a}}  & \colhead{$H_R$\tablenotemark{b}} & \colhead{B-V} & \colhead{V-R} & \colhead{B-R}  & \colhead{Source} }
\startdata

2014 QO441  & 	2016 Aug 03 & 23.00$\pm$0.01 & 7.6 	& 0.75$\pm$0.03 & 0.47$\pm$0.03 & 1.22$\pm$0.02 & This work \\
2011 SO277  &  	2016 Aug 03 & 22.54$\pm$0.03 & 7.5 	& 0.69$\pm$0.03 & 0.39$\pm$0.03 & 1.08$\pm$0.03 & This work \\
2013 KY18    &  	2016 Aug 03 & 21.29$\pm$0.01 & 6.3 	& 0.76$\pm$0.01 & 0.36$\pm$0.02 & 1.12$\pm$0.01 & This work \\
2011 WG157  &  	2016 Aug 03 & 21.95$\pm$0.04 & 6.8 	& 0.72$\pm$0.04 & 0.40$\pm$0.05 & 1.15$\pm$0.04 & This work \\
2010 TS191   &  	2016 Aug 03 & 22.39$\pm$0.03 & 7.6 	& 0.76$\pm$0.04 & 0.39$\pm$0.05 & 1.04$\pm$0.04 & This work \\
2010 TT191  &  	2016 Aug 03 & 22.74$\pm$0.03 & 7.4 	& 0.75$\pm$0.03 & 0.47$\pm$0.04 & 1.22$\pm$0.04 & This work \\
2011 HM102 & 2012 May 24  & 22.34$\pm$0.04 & 7.8 & 0.72$\pm$0.04 & 0.41$\pm$0.04 & 1.16$\pm$0.06 & Parker et al.~2013 \\
2007 VL305 & 2012 May 24  & 22.53$\pm$0.03 & 8.0 & 0.83$\pm$0.05 & 0.47$\pm$0.05 & 1.30$\pm$0.07 & Parker et al.~2013 \\
2006 RJ103 & 2012 May 24  & 21.80$\pm$0.04 & 6.9 & 0.82$\pm$0.03 & 0.47$\pm$0.03 & 1.29$\pm$0.04 & Parker et al.~2013 \\
2011 QR322 & 2004 - 2006 & 22.50$\pm$0.01 & 7.8 & 0.80$\pm$0.03 & 0.46$\pm$0.02 & 1.26$\pm$0.04 & ST06 \\
2004 UP10 & 2004 - 2006 & 23.28$\pm$0.03 & 8.5 & 0.74$\pm$0.05 & 0.42$\pm$0.04 & 1.16$\pm$0.07 & ST06 \\
2005 TN53 & 2004 - 2006 & 23.73$\pm$0.04 & 8.6 & 0.82$\pm$0.08 & 0.47$\pm$0.07 & 1.29$\pm$0.11 & ST06 \\
2005 TO74 & 2004 - 2006 & 23.21$\pm$0.03 & 8.1 & 0.85$\pm$0.06 & 0.49$\pm$0.05 & 1.34$\pm$0.08 & ST06 \\
\hline
Solar Colors & & & & 0.64$\pm$0.02 & 0.35$\pm$0.01 & 0.99$\pm$0.02 & Holmberg et al.~2006 \\

\enddata


\tablenotetext{a}{Mean apparent R magnitude and $\pm$1$\sigma$ uncertainty }
\tablenotetext{b}{R magnitude corrected to $r_H = \Delta$ = 1 AU and $\alpha$ = 0\degr. Values are quoted only to one decimal place in recognition of the unmeasured phase function, which introduces an uncertainty to $H_R$ of order 0.1 magnitude.}

\end{deluxetable}

\clearpage

\begin{deluxetable}{llcrrrccccr}
\tablecaption{Optical Colors\tablenotemark{a}
\label{colors}}
\tablewidth{0pt}
\tablehead{ \colhead{Group} &  \colhead{N\tablenotemark{b}}&  \colhead{B-V} & \colhead{V-R}  & \colhead{B-R} }
\startdata

NT & 13 & 0.77$\pm$0.01/0.76 & 0.44$\pm$0.01/0.46 & 1.20$\pm$0.03/1.22 \\
JT & 74 & 0.78$\pm$0.01/0.75 & 0.45$\pm$0.01/0.45 & 1.22$\pm$0.01/1.22 \\
SKBO & 53 & 0.89$\pm$0.02/0.86 & 0.54$\pm$0.01/0.53 & 1.42$\pm$0.03/1.39 \\
H-C & 41 & 0.93$\pm$0.03/0.93 & 0.57$\pm$0.02/0.59 & 1.50$\pm$0.04/1.53 \\
3:2 & 39 & 0.90$\pm$0.03/0.86 & 0.57$\pm$0.02/0.59 & 1.47$\pm$0.05/1.39 \\
Cen & 27 & 0.87$\pm$0.04/0.79 & 0.57$\pm$0.02/0.51 & 1.43$\pm$0.06/1.25 \\
C-C & 43 & 1.06$\pm$0.02/1.06 & 0.65$\pm$0.02/0.66 & 1.72$\pm$0.03/1.73 \\

\enddata


\tablenotetext{a}{For each group we list the mean, the standard error on the mean, and the median.}
\tablenotetext{b}{Number of objects in the group}

\end{deluxetable}


\clearpage

\begin{landscape}
\begin{deluxetable}{llcrrrccccr}
\tabletypesize{\scriptsize}

\tablecaption{Kolmogorov-Smirnov and Anderson-Darling Probabilities\tablenotemark{a}
\label{KS}}
\tablewidth{0pt}
\tablehead{ \colhead{Group\tablenotemark{b}} &  \colhead{NT}  & \colhead{JT}  & \colhead{H-C} & \colhead{SKBO} & \colhead{Cen} & \colhead{C-C} & \colhead{3:2}   }
\startdata
NT(13)    	& 1.000/1.000   &   0.839/0.714   & \textbf{0.003}/\textbf{$<$0.001} &    0.012/0.004 &  0.029/0.064 & \textbf{$<$0.001}/\textbf{$<$0.001} & \textbf{0.002}/\textbf{0.001} \\
JT(74)    	&  & 1.000/1.000 & \textbf{$<$0.001}/\textbf{$<$0.001} & \textbf{$<$0.001}/\textbf{$<$0.001} & \textbf{0.001}/\textbf{0.003} & \textbf{$<$0.001}/\textbf{$<$0.001} & \textbf{$<$0.001}/\textbf{$<$0.001} \\
H-C(41)  &  &  & 1.000/1.000 & 0.071/0.023   & 0.389/0.022   & \textbf{$<$0.001}/\textbf{$<$0.001} & 0.779/0.346\\
SKBO (53)   &    &   &   &  1.000/1.000  & 0.432/0.210  & \textbf{$<$0.001}/\textbf{$<$0.001} & 0.135/0.223\\
Cen (27)       &    &   &   &    &  1.000/1.000 & \textbf{0.001}/\textbf{$<$0.001} & 0.550/0.249\\
C-C (43)    &  &  &   &  &  & 1.000/1.000 & \textbf{0.001}/\textbf{$<$0.001}\\
3:2 (39)  &  &  &   &  &  &  & 1.000/1.000\\

\enddata

\tablenotetext{a}{Non-parametric probability that any two given color distributions could be drawn from the same parent population.  Results from the Kolmogorov-Smirnov and Anderson-Darling tests are written KS/AD.  Values with significance $P \le$ 0.003, indicating a small chance of being drawn from the same parent population,  are highlighted in \textbf{bold text}. Lower half of the diagonally symmetric matrix is not shown.}
\tablenotetext{b}{Numbers in parentheses give the sample size in each group}

\end{deluxetable}
\end{landscape}

\clearpage

\clearpage

\clearpage

\begin{thebibliography}{}

\bibitem[Alexandersen et al.(2016)]{2016AJ....152..111A} Alexandersen, M., Gladman, B., Kavelaars, J.~J., et al.\ 2016, \aj, 152, 111 


\bibitem[Anderson \& Darling(1954)]{1954AD} Anderson, T.W.; Darling, D.A.\ 1954. Journal of the American Statistical Association. 49, 765.


\bibitem[Benecchi \& Sheppard(2013)]{2013AJ....145..124B} Benecchi, S.~D., \& Sheppard, S.~S.\ 2013, \aj, 145, 124 


\bibitem[Bowell et al.(1989)]{1989The Book}  Bowell, E., Hapke, B., Domingue, D., et al. 1989, in Asteroids II, ed. R. Binzel,
T. Gehrels, \& S. Matthews, (Tucson, AZ: Univ. Arizona Press), 524

\bibitem[Chatelain et al.(2016)]{2016Icar..271..158C} Chatelain, J.~P., Henry, T.~J., French, L.~M., Winters, J.~G., \& Trilling, D.~E.\ 2016, \icarus, 271, 158 


\bibitem[Chiang \& Lithwick(2005)]{2005ApJ...628..520C} Chiang, E.~I., \& Lithwick, Y.\ 2005, \apj, 628, 520 

\bibitem[Cruikshank et al.(1998)]{1998Icar..135..389C} Cruikshank, D.~P., Roush, T.~L., Bartholomew, M.~J., et al.\ 1998, \icarus, 135, 389 

\bibitem[Dalle Ore et al.(2015)]{2015Icar..252..311D} Dalle Ore, C.~M., Barucci, M.~A., Emery, J.~P., et al.\ 2015, \icarus, 252, 311 

\bibitem[Dandy et al.(2003)]{2003Icar..163..363D} Dandy, C.~L., Fitzsimmons, A., \& Collander-Brown, S.~J.\ 2003, \icarus, 163, 363 

\bibitem[Duffard et al.(2009)]{2009A&A...505.1283D} Duffard, R., Ortiz, J.~L., Thirouin, A., Santos-Sanz, P., \& Morales, N.\ 2009, \aap, 505, 1283 

\bibitem[Fraser et al.(2008)]{2008Icar..195..827F} Fraser, W.~C., Kavelaars, J.~J., Holman, M.~J., et al.\ 2008, \icarus, 195, 827 

\bibitem[Fraser et al.(2014)]{2014ApJ...782..100F} Fraser, W.~C., Brown, M.~E., Morbidelli, A., Parker, A., \& Batygin, K.\ 2014, \apj, 782, 100 



\bibitem[Gomes \& Nesvorn{\'y}(2016)]{2016A&A...592A.146G} Gomes, R., \& Nesvorn{\'y}, D.\ 2016, \aap, 592, A146 

\bibitem[Grav et al.(2012)]{2012ApJ...759...49G} Grav, T., Mainzer, A.~K., Bauer, J.~M., Masiero, J.~R., \& Nugent, C.~R.\ 2012, \apj, 759, 49 

\bibitem[Hainaut et al.(2012)]{2012A&A...546A.115H} Hainaut, O.~R., Boehnhardt, H., \& Protopapa, S.\ 2012, \aap, 546, A115 

\bibitem[Holmberg et al.(2006)]{2006MNRAS.367..449H} Holmberg, J., Flynn, C., \& Portinari, L.\ 2006, \mnras, 367, 449 

\bibitem[Horner \& Lykawka(2010)]{2010MNRAS.402...13H} Horner, J., \& Lykawka, P.~S.\ 2010, \mnras, 402, 13 

\bibitem[Horner \& Lykawka(2012)]{2012MNRAS.426..159H} Horner, J., \& Lykawka, P.~S.\ 2012, \mnras, 426, 159 


\bibitem[Jewitt(2002)]{2002AJ....123.1039J} Jewitt, D.,\ 2002, \aj, 123, 1039 

\bibitem[Jewitt(2009)]{2009AJ....137.4296J} Jewitt, D.,\ 2009, \aj, 137, 4296 

\bibitem[Jewitt(2015)]{2015AJ....150..201J} Jewitt, D.,\ 2015, \aj, 150, 201 

\bibitem[Jewitt \& Luu(2001)]{2001AJ....122.2099J} Jewitt, D., \& Luu, J.,\ 2001, \aj, 122, 2099 


\bibitem[Kortenkamp et al.(2004)]{2004Icar..167..347K} Kortenkamp, S.~J., Malhotra, R., \& Michtchenko, T.\ 2004, \icarus, 167, 347 

\bibitem[Lacerda et al.(2014)]{2014ApJ...793L...2L} Lacerda, P., Fornasier, S., Lellouch, E., et al.\ 2014, \apjl, 793, L2 


\bibitem[Luu \& Jewitt(1996)]{1996AJ....112.2310L} Luu, J., \& Jewitt, D.\ 1996, \aj, 112, 2310 


\bibitem[Lykawka et al.(2009)]{2009MNRAS.398.1715L} Lykawka, P.~S., Horner, J., Jones, B.~W., \& Mukai, T.\ 2009, \mnras, 398, 1715 

\bibitem[Marzari \& Scholl(1998)]{1998A&A...339..278M} Marzari, F., \& Scholl, H.\ 1998, \aap, 339, 278 


\bibitem[Morbidelli et al.(2005)]{2005Natur.435..462M} Morbidelli, A., Levison, H.~F., Tsiganis, K., \& Gomes, R.\ 2005, \nat, 435, 462 

\bibitem[Morbidelli et al.(2009)]{2009Icar..202..310M} Morbidelli, A., Levison, H.~F., Bottke, W.~F., Dones, L., \& Nesvorn{\'y}, D.\ 2009, \icarus, 202, 310 

\bibitem[Oke et al.(1995)]{1995PASP..107..375O} Oke, J.~B., Cohen, J.~G., Carr, M., et al.\ 1995, \pasp, 107, 375 


\bibitem[Parker et al.(2013)]{2013AJ....145...96P} Parker, A.~H., Buie, M.~W., Osip, D.~J., et al.\ 2013, \aj, 145, 96 

\bibitem[Parker(2015)]{2015Icar..247..112P} Parker, A.~H.\ 2015, \icarus, 247, 112 

\bibitem[Peixinho et al.(2012)]{2012A&A...546A..86P} Peixinho, N., Delsanti, A., Guilbert-Lepoutre, A., Gafeira, R., \& Lacerda, P.\ 2012, \aap, 546, A86 

\bibitem[Peixinho et al.(2015)]{2015A&A...577A..35P} Peixinho, N., Delsanti, A., \& Doressoundiram, A.\ 2015, \aap, 577, A35 

\bibitem[Sheppard(2010)]{2010AJ....139.1394S} Sheppard, S.~S.\ 2010, \aj, 139, 1394 

\bibitem[Sheppard(2012)]{2012AJ....144..169S} Sheppard, S.~S.\ 2012, \aj, 144, 169 

\bibitem[Sheppard \& Trujillo(2006)]{2006Sci...313..511S} Sheppard, S.~S., \& Trujillo, C.~A.\ 2006, Science, 313, 511 

\bibitem[Shevchenko et al.(2012)]{2012Icar..217..202S} Shevchenko, V.~G., Belskaya, I.~N., Slyusarev, I.~G., et al.\ 2012, \icarus, 217, 202 

\bibitem[Slyusarev \& Belskaya(2014)]{2014SoSyR..48..139S} Slyusarev, I.~G., \& Belskaya, I.~N.\ 2014, Solar System Research, 48, 139 

\bibitem[Stephens \& Noll(2006)]{2006AJ....131.1142S} Stephens, D.~C., \& Noll, K.~S.\ 2006, \aj, 131, 1142 

\bibitem[Szab{\'o} et al.(2007)]{2007MNRAS.377.1393S} Szab{\'o}, G.~M., Ivezi{\'c}, {\v Z}., Juri{\'c}, M., \& Lupton, R.\ 2007, \mnras, 377, 1393 

\bibitem[Szab{\'o} et al.(2017)]{2017A&A...599A..44S} Szab{\'o}, G.~M., P{\'a}l, A., Kiss, C., et al.\ 2017, \aap, 599, A44 

\bibitem[Tegler \& Romanishin(2000)]{2000Natur.407..979T} Tegler, S.~C., \& Romanishin, W.\ 2000, \nat, 407, 979 

\bibitem[Thirouin et al.(2014)]{2014A&A...569A...3T} Thirouin, A., Noll, K.~S., Ortiz, J.~L., \& Morales, N.\ 2014, \aap, 569, A3 

\bibitem[Vinogradova \& Chernetenko(2015)]{2015SoSyR..49..391V} Vinogradova, T.~A., \& Chernetenko, Y.~A.\ 2015, Solar System Research, 49, 391 

\bibitem[Volk \& Malhotra(2008)]{2008ApJ...687..714V} Volk, K., \& Malhotra, R.\ 2008, \apj, 687, 714-725 

\bibitem[Wolff et al.(2012)]{2012ApJ...746..171W} Wolff, S., Dawson, R.~I., \& Murray-Clay, R.~A.\ 2012, \apj, 746, 171 

\bibitem[Wong \& Brown(2016)]{2016AJ....152...90W} Wong, I., \& Brown, M.~E.\ 2016, \aj, 152, 90 

\bibitem[Wong \& Brown(2017)]{2017AJ....153..145W} Wong, I., \& Brown, M.~E.\ 2017, \aj, 153, 145 

\bibitem[Yoshida \& Terai(2017)]{2017AJ....154...71Y} Yoshida, F., \& Terai, T.\ 2017, \aj, 154, 71 


\end{thebibliography}
\end{document}